\documentclass[%
reprint,
superscriptaddress,
 amsmath,amssymb,
 aps,
longbibliography
]{revtex4-1}

\usepackage{graphicx}
\usepackage{dcolumn}
\usepackage{bm}
\usepackage{xcolor}
\usepackage{hyperref}
\definecolor{bleuf}{rgb}{0,0.44,0.72}
\hypersetup{
    colorlinks=true,                          
    linkcolor=bleuf, 
    citecolor=bleuf, 
    urlcolor=bleuf  } 

\usepackage{placeins} 

\newcommand\grad\nabla

\begin{document}

\title{
Freezing a Flock:\\ 
Motility-Induced Phase Separation in Polar Active Liquids}

\author{Delphine Geyer}
\affiliation{Univ. Lyon, ENS de Lyon, Univ. Claude Bernard, CNRS, Laboratoire de Physique, F-69342, Lyon, France}
\author{David Martin}
\affiliation{Laboratoire Matiere et Systemes Complexes, UMR 7057 CNRS/P7, Universit\'e Paris Diderot, 10 rue Alice Domon et Leonie Duquet, 75205 Paris cedex 13, France}
\author{Julien Tailleur}
\affiliation{Laboratoire Matiere et Systemes Complexes, UMR 7057 CNRS/P7, Universit\'e Paris Diderot, 10 rue Alice Domon et Leonie Duquet, 75205 Paris cedex 13, France}
\author{Denis Bartolo}
\affiliation{Univ. Lyon, ENS de Lyon, Univ. Claude Bernard, CNRS, Laboratoire de Physique, F-69342, Lyon, France}
\date{\today}
\begin{abstract}
Combining model experiments and theory, we investigate the dense
phases of polar active matter beyond the conventional flocking
picture. We show that above a critical density flocks assembled from
self-propelled colloids arrest their collective motion, lose their
orientational order and form solids that actively rearrange their
local structure while continuously melting and freezing at their
boundaries.  We establish that active solidification is a first-order
dynamical transition: active solids nucleate, grow, and slowly coarsen
until complete phase separation with the polar liquids they coexists
with. We then theoretically elucidate this phase behaviour by
introducing a minimal hydrodynamic description of dense polar flocks
and show that the active solids originate from a Motility-Induced Phase
Separation. We argue that the suppression of collective motion in the
form of solid jams is a generic feature of flocks assembled from
motile units that reduce their speed as density increases, a feature
common to a broad class of active bodies, from synthetic colloids to
living creatures.
\end{abstract}
\maketitle
\section{Introduction}
The emergence of collective motion in groups of living creatures 
or synthetic motile units is now a well established physical process~\cite{Marchetti_review,Toner_review,Vicsek_Review,Cavagna_Review,Zottl2016,Granick}: Self-propelled particles move coherently  along the same direction whenever velocity-alignment interactions overcome orientational perturbations favoring isotropic random motion. 
This minimal picture goes back to Vicsek's seminal work~\cite{Vicsek1995} and  made it possible to elucidate the  flocking dynamics of systems as diverse as bird groups, polymers surfing on motility assays, shaken grains, active colloidal fluids and drone fleets~\cite{Bausch2010,Sumino2012,Bricard2013,Granick,Dauchot2010,Sood2014,Vasarhelyi2018}.
From a theoretical perspective, flocks are described as flying ferromagnets where point-wise spins move at constant speed along their spin direction~\cite{Vicsek1995,Vicsek_Review,Marchetti_review,Farrell2012,Gregoire2004}. However, this simplified description is inapt to capture the dynamics of dense populations where contact interactions interfere with self-propulsion and ultimately arrest the particle dynamics. Until now, aside from rare theoretical exceptions~\cite{Peruani2011PRL,Farrell2012,Peruani2013,Nesbitt2019,Sansa2018}, the consequences of motility reduction in dense flocks has remained virtually uncharted  despite its relevance to a spectrum of active bodies ranging from marching animals to robot fleets and active colloids. 

In this article, combining quantitative experiments and theory, we investigate the suppression of collective motion in high-density flocks.  We show and explain how polar assemblies of motile colloids turn into lively solid phases that actively rearrange their amorphous
structure, 
but do not support any directed motion. We establish that active solidification of polar liquids is a first-order dynamical transition:
active solids nucleate, grow, and slowly coarsen until complete phase separation. Even though they are mostly formed of particles at rest, we show that
active solids steadily propagate through the polar liquids they coexist with. Using numerical simulations and analytical theory, we 
elucidate all our experimental findings and demonstrate that the
solidification of colloidal flocks provides a realization of
the long sought-after 
complete Motility-Induced Phase Separation~\cite{Cates_Review,Fily2012PRL,Buttinoni2013,Redner2013PRL,Bialke2013EPL,Zottl2016,Liu2017}. 

\section{Solidification of colloidal flocks}
\label{sec:Solid}
\begin{figure*}
  \begin{center}
    \includegraphics[width=\textwidth]{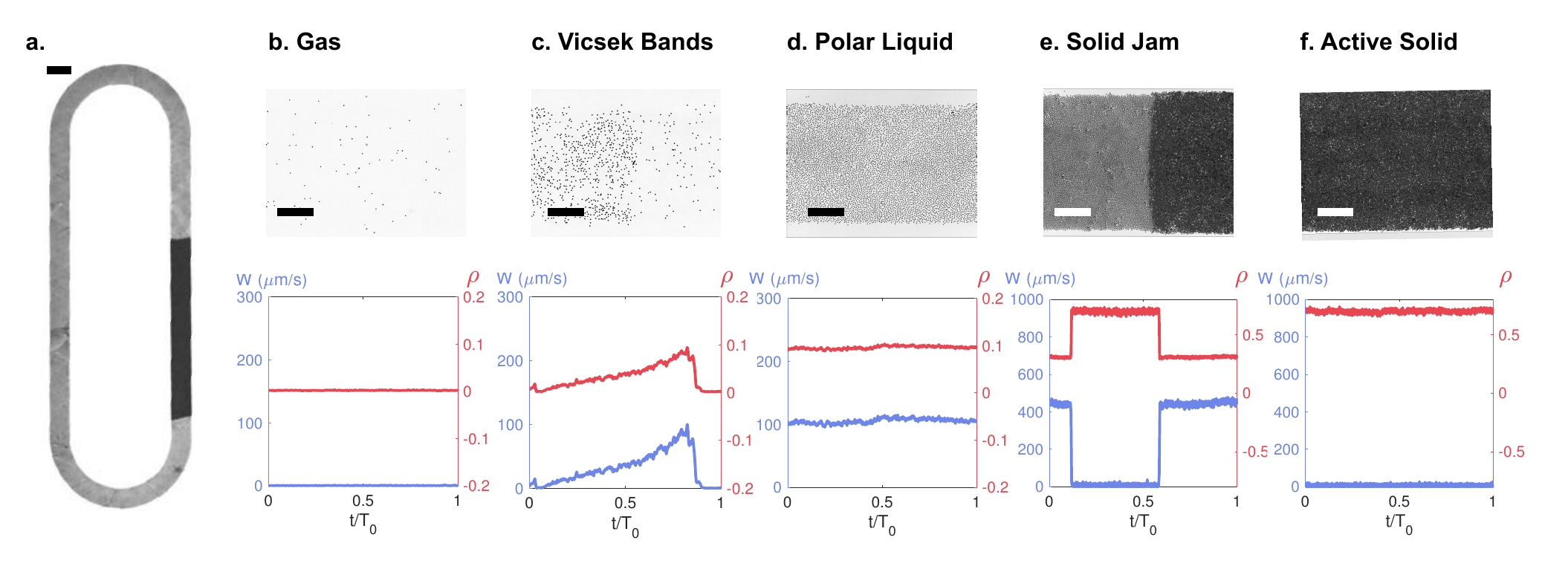}
    \caption{{\bf The dynamical phases of Quincke rollers.}
      \textbf{a}. Picture of a microfluidic racetrack where $\sim
      3\times$10$^6$ Quincke rollers interact.  An active solid (dark
      grey) propagates though a polar liquid (light grey). Scale bar:
      $2\,\rm mm$. \textbf{b--f.} Top panel: close-up pictures of
      Quincke rollers in the racetrack. Scale bars: $250\,\rm \mu
      m$. Bottom panel: longitudinal component of the particle current
      $W$ and density plotted as a function of a normalized time. Both
      $W$ and $\rho$ are averaged over an observation window of size
      $56\,\rm \mu m\times 1120\,\mu m$. $T_0$ is arbitrarily chosen to
      be the time taken by an active solid to circle around the race
      track. \textbf{b.} Gas phase ($\rho_0=0.002$). The density is
      homogeneous and the system does not support any net particle
      current. \textbf{c.} Coexistence between an active gas and a
      denser polar band ($\rho_0=0.033$). A heterogeneous polar-liquid
      drop propagates at constant speed through a homogeneous
      isotropic gas in the form of a so-called Vicsek band. \textbf{d.}
      Polar-liquid phase ($\rho_0=0.096$). The homogeneous active
      fluid supports a net flow. \textbf{e.} Coexistence between a
      polar liquid and an amorphous active solid ($\rho_0=0.49$). The
      high-density solid is homogeneous but, unlike the polar liquid,
      does not support any net particle current. \textbf{f.}
      Homogeneous active solid phase ($\rho_0=0.70$). }
    \label{Fig1}
  \end{center}
\end{figure*}
Our experiments are based on colloidal rollers. In brief, taking
advantage of the so-called Quincke
instability~\cite{Quincke,Bricard2013}, we motorize inert colloidal
beads and turn them into self-propelled rollers all moving at the same
constant speed $\nu_0=1040\,\rm \mu m/s$ when isolated, see also
Appendix~\ref{Method}. We observe $5\,\rm \mu m$ colloidal rollers
propelling and interacting in microfluidic racetracks of length
$L_0=9.8\,\rm cm$ and width $2\,\rm mm$, Fig.~\ref{Fig1}a. Both
hydrodynamic and electrostatic interactions promote alignment of the
roller velocities. As a result, the low-density phase behavior of the
Quincke rollers falls in the universality class of the Vicsek
model~\cite{Vicsek1995,Chate2008,Bertin2009JPA,Solon2015}. At low
packing fraction, Fig.~\ref{Fig1}b, they form an isotropic gas where
the density $\rho(\mathbf r,t)$ is homogeneous and the local particle
current $\mathbf \mathbf W(\mathbf r,t)$ vanishes. Increasing the
average packing fraction $\rho_0$ above $\rho_{\rm F}=0.02$,
Fig.~\ref{Fig1}c, the rollers undergo a flocking transition. The
transition is first-order, and polar liquid bands, where all colloids
propel on average along the same direction, coexist with an isotropic
gas~\cite{Bricard2013,Morin2017}.   Further increasing $\rho_0$,
the ordered phase fills the entire system and forms a homogeneous
polar liquid which flows steadily and uniformly as illustrated in
Supplementary Movie 1.  In polar liquids both $\mathbf W(\mathbf r,t)$
and $\rho(\mathbf r,t)$ display small (yet anomalous) fluctuations,
and orientational order almost saturates, see~Fig.~\ref{Fig1}d
and~\cite{Geyer2018}. This low-density behavior provides a
prototypical example of flocking physics.

\begin{figure} 
  \begin{center}
    \includegraphics[width=0.5\textwidth]{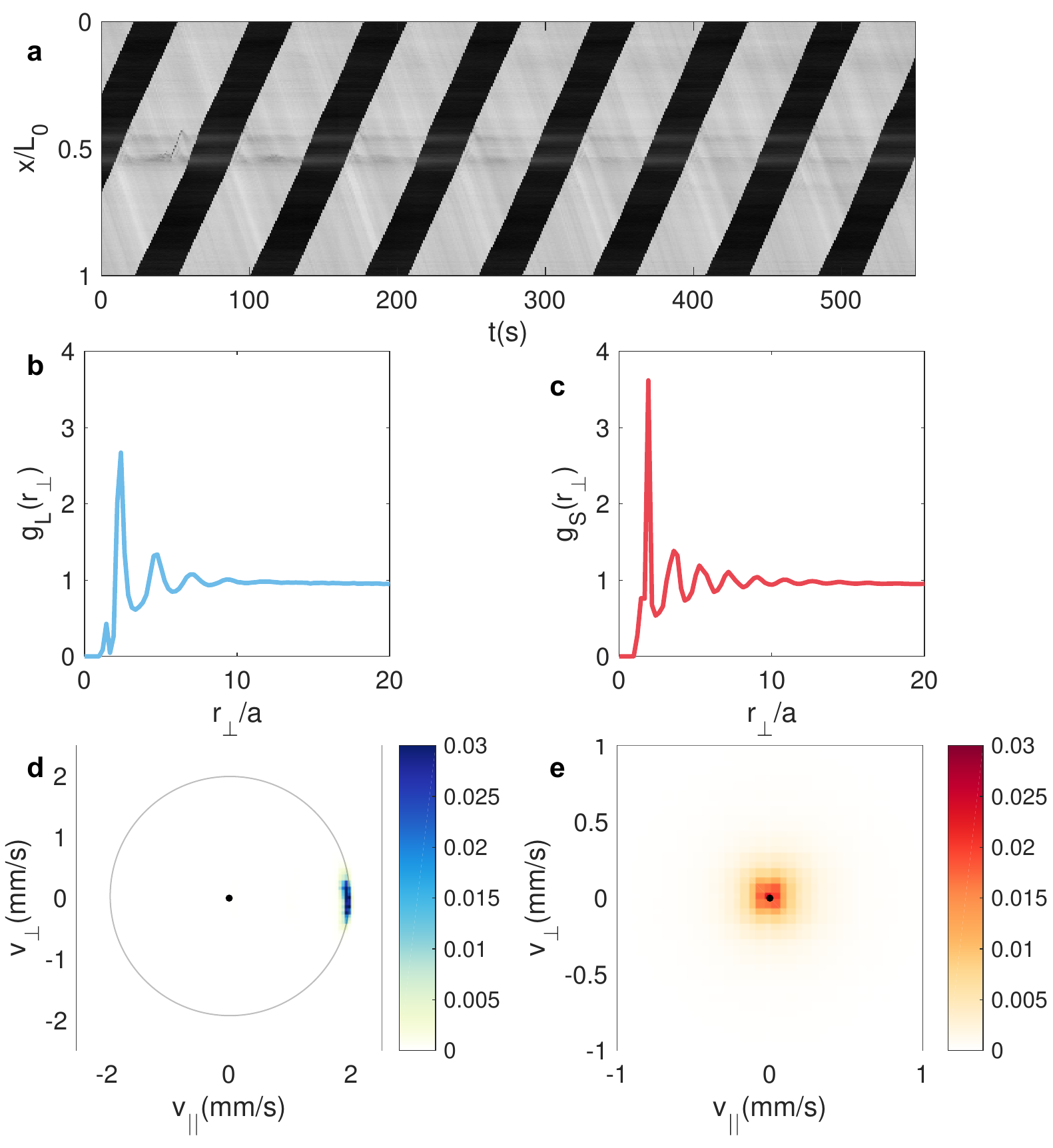}
    \caption{ \textbf{Structure and dynamics of active jams.}
      \textbf{a.} The kymograph of the measured light intensity
      averaged over the racetrack width shows how an active solid
      (dark region) propagates at constant speed through a homogeneous
      polar liquid (light region). \textbf{b and c.} Pair-correlation
      functions measured in the polar liquid ($g_{\rm L}$) and in the
      coexisting active-solid phase ($g_{\rm S}$). $\rho_0=0.58$. Both
      pair-correlation functions are plotted versus the interparticle
      distance $r_{\perp}$ in the direction transverse to the mean
      polar-liquid flow. $g_{\rm S}$ displays more peaks than $g_{\rm
        L}$ revealing a more ordered structure, but translational
      order merely persists over a few particle diameters.
      \textbf{d}. Probability density functions of the roller
      velocities in the polar liquid region.  The distribution is
      peaked around $\nu_0{ \hat{\mathbf x}_\parallel}$, where
      ${\hat{\mathbf x}_\parallel}$ is the vector tangent to the
      racetrack centerline. \textbf{e}. Probability density functions
      of the roller velocities in the active jam region. The
      distribution is peaked around 0. The rollers remain mostly at
      rest. \textbf{b}, \textbf{c}, \textbf{d}, and \textbf{e}:
      average area fraction $\rho_0=0.58$.}
    \label{Fig2}
  \end{center}
\end{figure}

However, when $\rho_0$ exceeds $\rho_{S}\simeq 0.55$ collective motion is
locally suppressed and flocking physics fails in explaining our
experimental observations. Particles stop their collective motion and
jam as exemplified in Supplementary Movies 2 and 3. 
The jams are active solids that continuously melt at one end while
growing at the other end. This lively dynamics hence preserves the
shape and length ($L_{\rm S}$) of the solid which propagates at
constant speed upstream the polar-liquid flow, see the kymograph of
Fig.~\ref{Fig2}a. Further increasing $\rho_0$, the solid region grows
and eventually spans the entire system, Fig.~\ref{Fig1}f.

Active solids form an amorphous phase. The pair correlation function
shown in Figs.~\ref{Fig2}b and \ref{Fig2}c indicate that active solids
are more spatially ordered than the polar liquid they coexist with,
but do not display any sign of long-range translational order. As
clearly seen in Supplementary Movie 2, the colloid dynamics are
however markedly different in the two phases. While they continuously
move at constant speed in the polar liquid, in the active solid, the
rollers spend most of their time at rest thereby suppressing any form
of collective motion and orientational order, Figs.~\ref{Fig2}d and
~\ref{Fig2}e.

We stress that the onset of active solidification corresponds to an area
fraction $\rho_{\rm S}\simeq 0.5$ which is much smaller than the random close
packing limit ($\rho_0=0.84$) and than the crystalization point of
self-propelled hard disks reported by Briand and Dauchot
($\rho_0\sim0.7$) in~\cite{Briand2016}. This marked difference hints
towards different physics which we characterize and elucidate below.
\section{The emergence of amorphous active solids is a first-order phase separation}

\begin{figure*} 
	\begin{center}
		\includegraphics[width=0.8\textwidth]{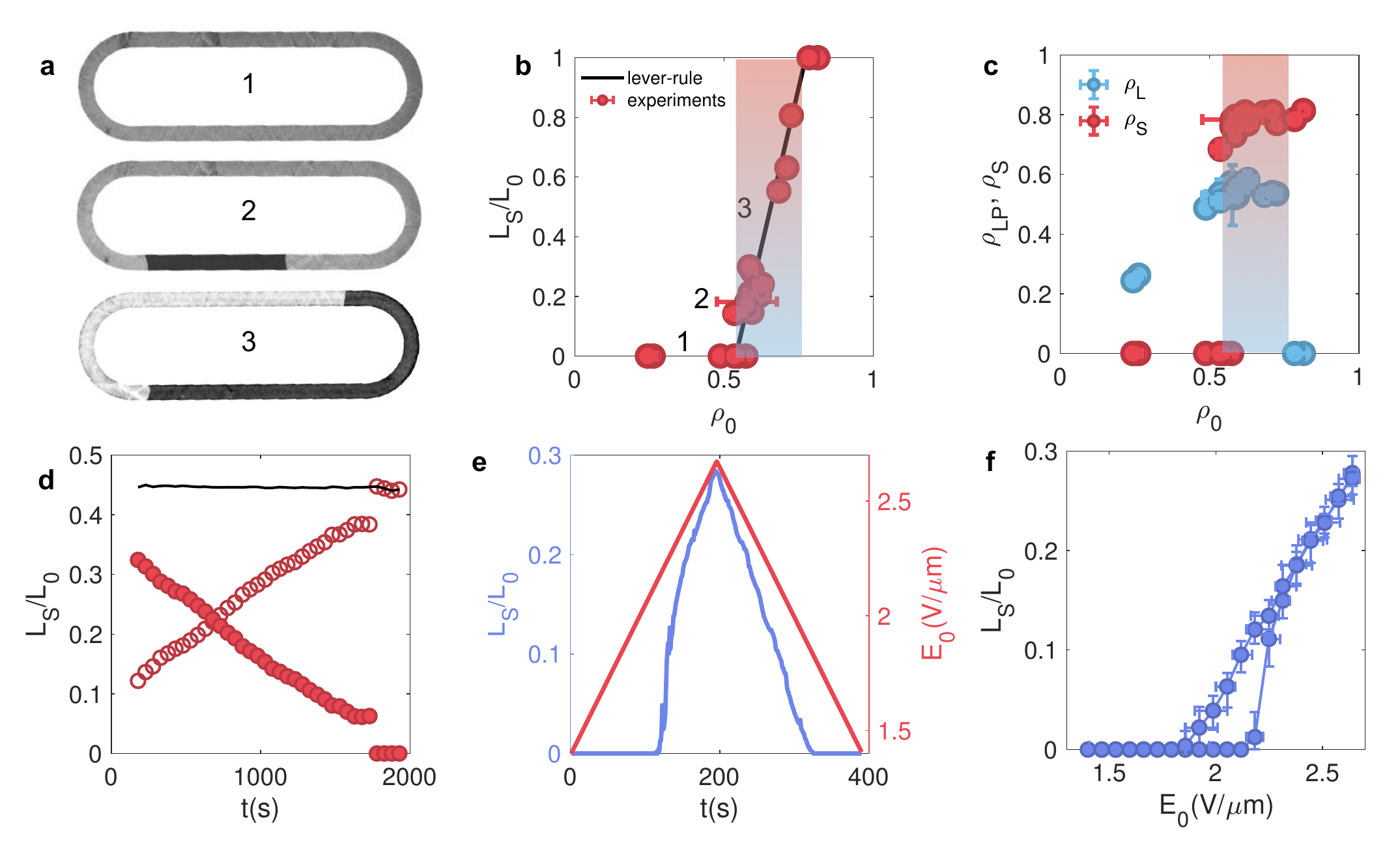}
		\caption{ \textbf{Active solidification is a first-order phase separation.} \textbf{a.} Solid jams in a racetrack at $\rho_0=0.38,\,0.58,\,0.65$ Increasing the average packing fraction, the extent of the active solid increases. \textbf{b.} Solid fraction plotted versus the average density $\rho_0$. Note the discontinuous jump and the two possible states at the onset of solidification. \textbf{c.} Density of the polar-liquid phase (blue circles) and of the active-solid phase (red circles) plotted versus $\rho_0$. In steady state,  the liquid density  increases with $\rho_0$ until an active solid forms, the density in both phases then remains constant. In {\textbf b.} and {\textbf c.} the shaded regions indicate the coexistence  between the polar liquid and an active solid  phases. \textbf{d.} Coarsening dynamics. Two solid jams coexist only over a finite time period. The larger jam (filled symbols)  shrinks and eventually vanishes in favor of the smaller one (open symbols).   The total fraction of  solid phase in the racetrack (black line) remains constant over time. \textbf{e.} Solid fraction (blue line) and magnitude of the electric field (red line) plotted versus  time. Over a range of  $E_0$ values, the active-solid fraction is different when increasing or decreasing the electric field. \textbf{f.} The extent of the traffic jam follows a hysteresis loop when cycling the field amplitude. } 
		\label{Fig3}
		\end{center}
\end{figure*}

We now establish that the formation of active solids occurs according
to a first order phase-separation scenario.  Firstly,
Figs.~\ref{Fig3}a and b indicate that, upon increasing $\rho_0$, the
extent of the solid phase has a lower bound: starting from a
homogeneous polar liquid, the solid length $L_{\rm S}$ discontinuously
jumps to a finite value before increasing linearly with
$\rho_0-\rho_{L}^{\rm b}$, where $\rho_{L}^{\rm b}=0.53$, see
Fig.\ref{Fig3}b.   The smallest solid
observed in a stationary state is as large as $L_{\rm S}\sim 1.4\rm
cm$. Smaller transient solid jams do form at local heterogeneities,
but they merely propagate over a finite distance before rapidly
melting, see Supplementary Movie 4. This observation suggests the
existence of a critical nucleation radius for the solid phase.

Secondly, while the packing fraction of a polar liquid obviously
increases with $\rho_0$, its value saturates as it coexists with an
active solid, Fig.~\ref{Fig3}c. At coexistence, the local densities in
the bulk of the liquid and solid phases are independent of the average
density $\rho_0$: $\rho_S^{\rm b}=0.77$ and $\rho_{L}^{\rm b}=0.53$,
which again supports a nucleation and growth scenario.  Increasing
$\rho_0$ leaves the inner structure of both phases unchanged and
solely increases the fraction of solid $L_{\rm S}/L_0$ in the
racetrack. We find that, as in equilibrium phase separation, the length
of the solid region is accurately predicted using a lever rule
constructed from the stationary bulk densities $\rho_{\rm S}^{\rm b}$
and $\rho_{\rm L}^{\rm b}$, see Fig.~\ref{Fig3}b.

Thirdly, we stress that when multiple jams nucleate in the device,
they propagate nearly at the same speed, see Supplementary Movie
4. Therefore, they cannot catch up and coalesce. The system in fact
reaches a stationary state thanks to a slow coarsening dynamics
illustrated in Fig.~\ref{Fig3}d where we show the temporal evolution
of the length of two macroscopic active solids (red symbols) and of
the overall solid fraction (dark line). One solid jam grows at the
expense of the other and coarsening operates leaving the overall
fraction of solid constant.  All of our experiments end with complete
phase separation: a single macroscopic active solid coexists with a
single active liquid phase. The final state of the system is therefore
uniquely determined by two macroscopic control parameters: the average
density $\rho_0$ and the magnitude $E_0$ of the electric field used to
power the rollers.

Finally, the most compelling argument in favour of a genuine
first-order phase separation is the bistability of the
two phases. Fig.~\ref{Fig3}b shows that at the onset of
solidification, depending on the (uncontrolled) initial conditions,
the system is either observed in a homogeneous polar liquid or at
liquid-solid coexistence. The bistability of the active material is
even better evidenced when cycling the magnitude of $E_0$ (cycling the
average density is not experimentally feasible). Fig.~\ref{Fig3}e
shows the temporal variations of the active-solid fraction upon 
triangular modulation of $E_0$, see also Supplementary Movie 5. When
$E_0$ increases an active-solid nucleates and quickly grows. When $E_0$
decreases, the solid continuously shrinks and eventually vanishes at a
field value smaller than the nucleation point. The asymmetric dynamics of $L_{\rm S}/L_0$ demonstrates the
existence of a metastable region in the phase diagram.  As shown in
Fig.~\ref{Fig3}f, the metastability of the active solid results in the hysteretic response of $L_{\rm S}$,  the hallmark of a first-order phase transition. We also note that the
continuous interfacial melting  observed when $E_0$  smoothly decreases contrasts with the response to a rapid field
quench, see Supplementary Movie 6. Starting with a stationary active
solid, a rapid quench results in
a destabilization of the solid bulk akin to a spinodal decomposition dynamics.

Altogether these measurements firmly establish that the emergence of  active solids results from a first-order phase separation, which we theoretically elucidate below.
\section{Motility-Induced phase separation in high-density polar flocks}
\subsection{Nonlinear hydrodynamic theory}
\label{Sec:hydro}
\begin{figure*}
  \begin{center}
    \includegraphics[width=\textwidth]{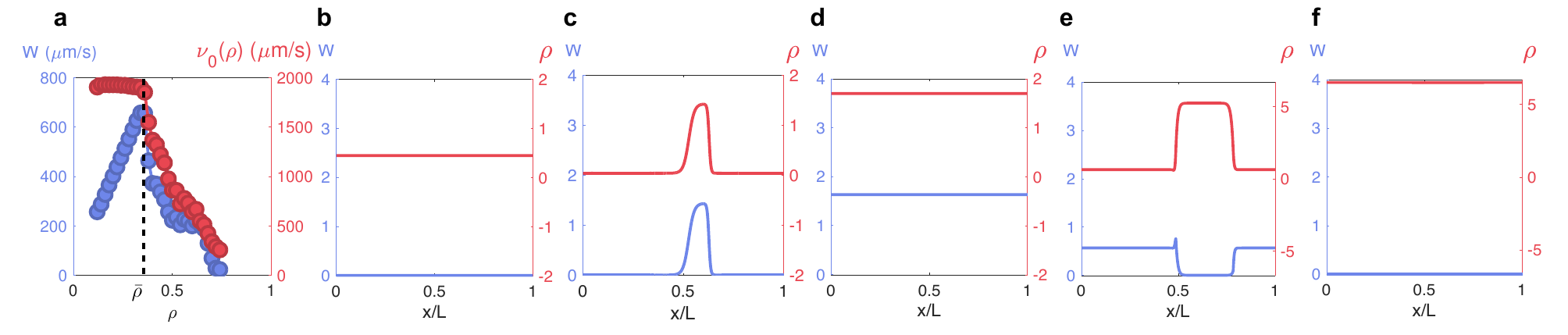}
    \caption{\textbf{Nonlinear hydrodynamics of polar active matter. }\textbf{a.} Average velocity $\nu_{0}(\mathbf r,t)$ of the colloids and local magnitude of the longitudinal current $W(\mathbf r,t)$ plotted as
      a function of the local density $\rho(\mathbf r,t)$. \textbf{b-e.}
      Successive phases observed in the resolution of
      Eqs.~\eqref{Toner-1} and~\eqref{Toner-2} at increasing densities
      $\rho = (0.10, 0.18, 1.7 ,1.96, 6.5)$. The position $x$ is normalized by the system size $L$. Simulation parameters: $D_{\rho}=0.2$, $D_{W}=1$,	$\lambda=1$, $a_{4}=0.45$, $\sigma_{1}=0.2$, $\bar{\rho}=2$,	$\xi=0.01$,	$\sigma_{2}=1$,  $\phi=0.5$, $L=200$,	 $dx=0.05$, $dt=0.005$.}
    \label{Fig4}
  \end{center}
\end{figure*}

As a last experimental result, we show in Fig.~\ref{Fig4}a how the
roller speed $\nu_0(\rho)$ varies with the {\em local} density
$\rho(\mathbf r,t)$ evaluated in square regions of size $12 a\sim
29\,\rm \mu m$. These measurements correspond to an experiment where a
solid jam coexists with a homogeneous polar liquid. $\nu_0(\rho)$
hardly varies at the smallest densities and sharply drops towards
$\nu_0(\rho)=0$ when $\rho(\mathbf r,t)$ exceeds $\bar \rho\sim0.35$.
Although we cannot positively identify
the microscopic origin of this abrupt slowing down, we detail
a possible explanation in Appendix~\ref{Quincke}. Simply put, the
lubrication interactions between nearby colloids with colinear
polarizations result in the reduction of their rotation rate which
ultimately vanishes at contact: near-field hydrodynamic interactions
frustrate self-propulsion. Instead of elaborating a microscopic
theory specific to colloidal rollers as
in~\cite{Bricard2013}, we instead adopt a generic hydrodynamic
description to account for all our experimental findings.

We start with a minimal version of the Toner-Tu equations which proved to correctly capture the coexistence of active gas and polar-liquid drops  at the onset of collective motion~\cite{Toner_review,Caussin2014PRL,Solon2015PRE}. For sake of simplicity we ignore fluctuations transverse to the mean-flow direction and write the hydrodynamic equations for the one-dimensional density $\rho(x,t)$ and longitudinal current $W(x,t)$:
\begin{align}
  \label{Toner-1}
  \partial_t\rho+\partial_x W&=D_\rho\partial_{xx} \rho,\\
  \label{Toner-2}
  \partial_t W+\lambda W \partial_x W&= D_{W}\partial_{xx} W - \partial_x \left[\epsilon_{1}(\rho)\rho\right]\nonumber
  \\  &+\left[\rho\epsilon_{2}(\rho)-\phi\right]W -a_{4}W^3.
\end{align}
We modify  the Toner-Tu hydrodynamics to account for the slowing down of the rollers  when $\rho$ exceeds $\bar \rho$,  Fig.~\ref{Fig4}a. Two terms must be modified to capture this additional physics:   the so-called pressure term $\rho\epsilon_1(\rho)$, and the
density-dependent alignment term $\rho\epsilon_2(\rho)$ responsible for the emergence of orientational order and collective motion.

Coarse-graining microscopic flocking models typically leads to
a pressure term proportional to the particle
speed~\cite{Bertin2009JPA,Marchetti_review}. We therefore expect
$\epsilon_1(\rho)$ to sharply decrease when $\rho(x,t)>\bar \rho$. At even higher densities, we also expect the repulsion and contact interactions between the particles to
result in a pressure increase with the particle
density~\cite{Bricard2013,Solon2015PRL}.  We henceforth disregard
this second regime which is not essential to 
the nucleation and  propagation of active solids.  The functional forms of $\epsilon_2(\rho)$ is phenomenologically deduced from the loss of orientational order in the solid phase  reported in Figs.~\ref{Fig2}e.  
This property is modelled by a function $\epsilon_2(\rho)$ which decreases from a constant positive value in the low-density phases to a vanishing value  deep in the solid phase. In all that follows,  we conveniently assume $\epsilon_2(\rho)$ and  $\epsilon_1(\rho)$ to be proportional. This assumption is supported by the  similar variations observed for the roller speed and  local current in Fig.~\ref{Fig4}a.
In
practice, we take $\epsilon_i=\sigma_i[1-\tanh((\rho-\bar
  \rho)/\xi)]$, where $\sigma_1$ and $\sigma_2$ are constant.

\begin{figure*} 
	\begin{center}
		\includegraphics[width=\textwidth]{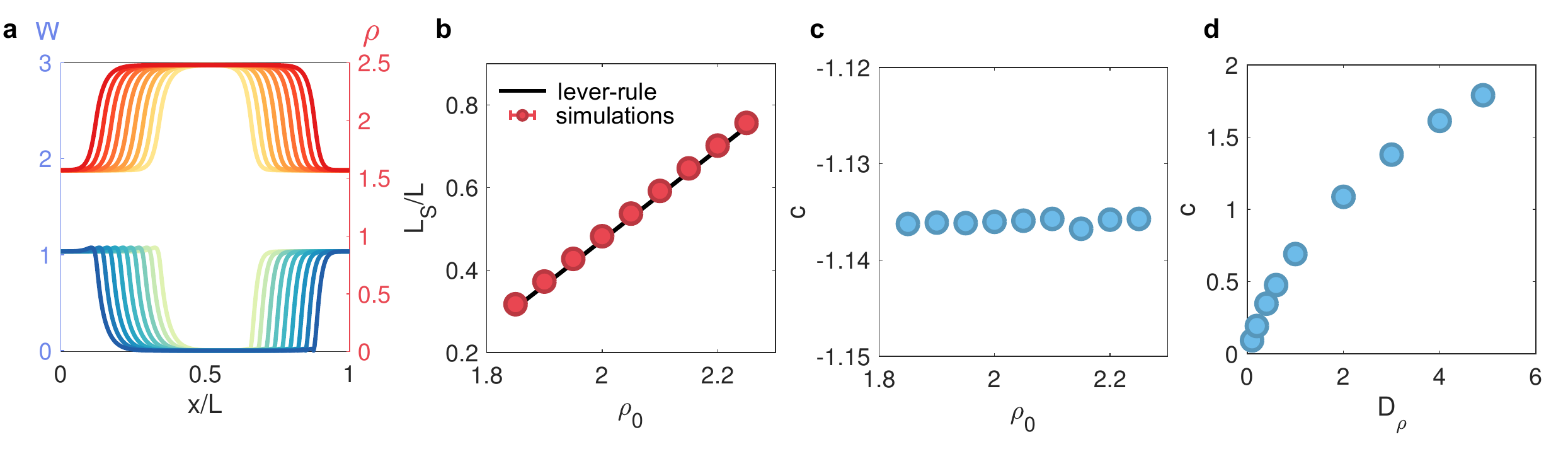}
 		\caption{\textbf{Shape and dynamics of active-solid jams.}  \textbf{a.} Density and velocity
                  profiles computed for different value of $\rho_0$ : from 1.85 (light colors) to 2.25 (dark colors). Numerical resolution of Eqs.~\eqref{Toner-1} and~\eqref{Toner-2} with the hydrodynamic parameters: $D_{\rho}=5$, $D_{W}=10$,	$\lambda=1$, $a_{4}=1$, $\sigma_{1}=1$, $\bar{\rho}=2$,	$\xi=0.1$,	$\sigma_{2}=1$,
$phi=0.5$, $L=200$,	 $dx=0.01$, $dt=0.001$. \textbf{b.}
                  Length of the solid jam  plotted versus $\rho_0$ (symbols) and lever rule (solid line). Same parameters as in {\textbf a.} \textbf{c.}
                  The speed of the solid jam $c$ does not depend on the average density $\rho_0$. \textbf{d.} Variations of the propagation speed $c$ as a function of the effective diffusivity $D_\rho$. Numerical parameters: $D_{W}=5$, $xi=1$, $a_{4}=0.45$, $\sigma_{1}=0.2$, $\bar{\rho}=2$, $\xi=0.01$,	$\sigma_{2}=1$, 
$phi=0.5$, $L=200$,	 $dx=0.05$, $dt=0.005$, $\rho_{0}=1.96$.}
		\label{Fig5}
		\end{center}
\end{figure*}

Numerical resolutions of Eqs.~\eqref{Toner-1} and \eqref{Toner-2} at
increasing densities faithfully account for the five successive phases observed in our
experiments, see Fig.~\ref{Fig4} and Appendix~\ref{Method}. At low densities, we first observe the
standard Vicsek transition: a
disordered gas phase is separated from a homogeneous polar liquid phase by a
coexistence region where ordered bands propagate through a disordered
background~\cite{Gregoire2004}. This phase transition occurs at very low area fraction ($\rho_0\sim\phi\ll\bar \rho$), in a regime where the colloidal rollers experience no form of kinetic hindrance as they interact, therefore $\epsilon_i(\rho)\simeq \sigma_i$. In agreement with 
our experiments, a second transition leads to the coexistence
between a polar liquid of constant density $\rho_L^{\rm b}$ and an apolar
dense phase of constant density $\rho_S^{\rm b}$. This jammed phase  propagates backwards with
respect to the flow of the polar liquid as does the active solids we observe in our experiments. This second transition shares all
the signatures of the first-order phase separation reported in Fig.~\ref{Fig3}. Figs.~\ref{Fig5}a, ~\ref{Fig5}b and ~\ref{Fig5}c indicate that the jammed phase obeys a
lever rule, its width increases linearly with $\rho_0-\rho_{L}^{\rm b}$,
while the velocity $c$ and the shape of the fronts remains
unchanged upon increasing $\rho_0$. 
 The first order nature of the transition is further supported by Supplementary Movie 7 which shows the existence of a
hysteresis loop when ramping up and down the average density. 
\subsection{Spinodal instability of polar liquids and domain wall propagation}
Having established the predictive power of our hydrodynamic model, we now use it to gain  physical insight into the origin of  active solidification.
We focus on the experimentally relevant situation where $\bar \rho\gg\phi_{\rm g}$, i.e. where the slowing down of the particle occurs at area fractions  much larger than the onset of collective motion. Given this hierarchy, at low densities, when $\rho_0\ll\bar \rho$, the hydrodynamic equations Eqs.~\eqref{Toner-1}-\eqref{Toner-2} correspond to that thoroughly studied in~\cite{Caussin2014PRL,Solon2015PRE,Baskaran2010}. They correctly predict a first order transition from an isotropic gas to a polar-liquid phase, see also Appendix~\ref{app:linstab}.  

The phase separation between a polar liquid and a jammed phase becomes
also clear when performing a linear stability analysis of the
homogeneous solutions of Eqs.~\eqref{Toner-1}-\eqref{Toner-2}, see
Appendix~\ref{app:linstab} where the stability of the various phases
is carefully discussed.  At high density, the stability of polar
liquids where $\rho=\rho_0$ and $W=W_0\neq 0$ is limited by a
phenomenon that is not captured by classical flocking models.  When
$\rho_0 \gg \phi + \epsilon_{2}/(2 \lambda \epsilon_{2}+ 4 a_4
\epsilon_1)$, polar liquids are stable with respect to the spinodal
decomposition into Vicsek bands, however another instability sets in
whenever $\epsilon_i'(\rho_0)\rho_0+ \epsilon_i(\rho_0)$ is
sufficiently large and negative, which occurs when $\rho$ approaches
$\bar \rho$.  This instability is responsible for the formation of
active-solid jams. We learn from the stability analysis that it
ultimately relies on the decrease of the effective pressure with
density in Eq.~\eqref{Toner-2} as a result of the slowing down of the
colloids in dense environments. This criterion is exactly analogous to
the spinodal decomposition condition in MIPS physics: the formation of
active-solid jams results from a complete motility induced phase
separation~\cite{Cates_Review}.

\subsection{Discussion}
Two comments are in order. Firstly, our results provide a novel
microscopic mechanism leading to MIPS. In classical systems such as
Active Brownian Particles, repulsive interactions and persistent
motion conspire to reduce the local current when active particles
undergo head-on
collisions~\cite{Fily2012PRL,Redner2013PRL,Bialke2013EPL}. Here we
show that this microscopic dynamics is however not necessary to
observe phase separation and MIPS transitions solely rely on the
reduction of the active particle current as density becomes
sufficiently high, irrespective of its microscopic origin. In the case
of colloidal rollers, particle indeed do not experience any frontal
collision when an active solid nucleate in a polar liquid, phase
separation is however made possible by the slowing down of their
rolling motion.

Secondly, another marked difference with the dense phases of
conventional MIPS system is the steady propagation of the active
solids through the dilute polar liquid.  This dynamics can be
accounted for by our model. The two boundaries of the active solid are
two domain walls that propagate at the same speed. The propagation of
the domain wall at the front of the solid jam relies on a mechanisms
akin to actual traffic jam propagation: the directed motion of the
particles incoming from the polar liquid cause an accumulation at the
boundary with the arrested phase, in the direction opposing the
spontaneous flow. By contrast, the propagation of the second domain
wall, at the back of the solid jam, requires arrested particles to
resume their motion. The formation of this smooth front originates
from the mass diffusion terms in Eq.~\eqref{Toner-1}, which allows
particles to escape the arrested solid phase and progressively resume
their collective motion when reaching a region of sufficiently low
density in the polar liquid. This diffusive spreading, however, does
not rely on thermal diffusion. The roller difusivity $D_{\rm m}\sim
10^{-13}\,\rm m^2/s$ is indeed negligible on the timescale of the
experiments.  Fortunately, other microscopic mechanisms, and in
particular anisotropic interactions, lead to diffusive contributions
to the density current~\cite{solon2015EPJST}.
A simple way to
model this effect is to consider a 
velocity-density relation of the form $\nu_0(\rho)[1 - r {\bf u} \cdot
  \grad \rho]$ where
${\bf u}$ is the orientation of the particle and $r$, which could be
density dependent, quantifies the anisotropic slowing down of  particles ascending density gradients. This anisotropic form is consistent with the polar symmetry of the flow and electric field induced by the Quincke rotation of the colloids. Coarse-graining the dynamics of
self-propelled particles interacting via such a nonlocal quorum sensing rule was
 done in~\cite{solon2015EPJST} and leads to an effective Fickian contribution $\sim-\rho \nu_0 r \grad \rho$ to the density current in Eq.~\eqref{Toner-1}. The ratio  between the magnitude of this effective Fickian flux and that of thermal diffusion  is readily estimated as $(\rho \nu_0 r)/D_{ m}\sim 10^5 $, assuming that $r$ is of the order of the a colloid diameter. Anisotropic interactions are therefore expected to strongly amplify the magnitude of $D_\rho$. In order to confirm the prominent role of this diffusion term in the active solid dynamics, we numerically measure the propagation speed $c$ of the jammed region as a function
of $D_\rho$. In agreement with the above discussion $c$ is found to vanish as $D_\rho \to 0$,
Fig.~\ref{Fig5}d thereby confirming the requirement of a finite diffusivity to observe stable active-solid jams. Simply put, the steady propagation of  active solids relies on the balance between two distinct macroscopic phenomena: motility reduction at high density which results in the formation of sharp interfaces with the polar liquid, and the diffusive smoothing of the interfaces that enables particles trapped in the arrested solid phase to resume their motion by rejoining the polar-liquid flock.    

\section{Conclusion}
In summary, combining expriments on Quincke rollers and active-matter
theory, we have shown that the phase behavior of polar active units is
controled by a series of two dynamical transitions: a Flocking
transition that transforms active gases into spontaneously flowing
liquids, and a Motility-Induced Phase Separation that results in the
freezing of these polar fluids and the formation of active
solids. Although most of their constituents are immobile, active solid
jams steadily propagate through the active liquid they coexist with
due to their continuous melting and freezing at their
boundaries. Remarkably, Quincke rollers provide a rare example of an
unhindered MIPS dynamics that is not bound to form only finite-size
clusters, see~\cite{Tjhung2018} and references therein. Beyond the
specifics of active rollers, we have shown that the freezing of
flocking motion and the emergence of active solids is a generic
feature that solely relies on polar ordering and speed reduction in
dense environements. A natural question to ask is wether suitably
tailored polar and quorum-sensing interactions could yield ordered
active solids. More generally understanding the inner structure and
dynamic of active solids is an open challenge to active matter
phycisists.

  \begin{acknowledgments}
D.~G. and D.~M. have equal contributions. We thank  A. Morin for invaluable help with the experiments.\\
\end{acknowledgments}

\appendix
\section{Quincke rollers.}
\label{Quincke}
\subsection{Motorization}
Our experiments are based on colloidal rollers~\cite{Bricard2013}.  We motorize inert polystyrene colloids of radius $a=2.4\,\rm\mu m$ by taking advantage of the so-called Quincke electro-rotation instability~\cite{Quincke,Taylor69}. Applying a DC electric field to an insulating body immersed in a conducting fluid results in  a  surface-charge dipole $\mathbf P$.  Increasing the magnitude of the electric field $\mathbf E_0$ above the Quincke threshold $E_{\rm Q}$  destabilizes  the dipole orientation, which in turn makes a finite angle with the electric field. A net electric torque $\mathbf T_{\rm E}\sim \mathbf P\times \mathbf E_0$ builds up and competes with a viscous frictional torque $\mathbf T_{\rm V}\sim\eta \mathbf\Omega$ where $\mathbf \Omega$ is the colloid rotation rate and   $\eta$ is the fluid shear viscosity. In steady state,  the two torques balance and the colloids rotate at constant angular velocity. As sketched in Fig.~\ref{FigSupp1}a and ~\ref{FigSupp1}b, when the colloids are let to sediment on a flat electrode, rotation is  readily converted into translationnal motion at constant speed $v_0$ (in the direction opposite to the charge dipole).
We stress that the direction of motion is randomly chosen and freely diffuses as a result of the spontaneous symmetry breaking of the surface-charge distribution.

\subsection{Arresting Quincke rotation}
\begin{figure} 
	\begin{center}
		\includegraphics[width=0.5\textwidth]{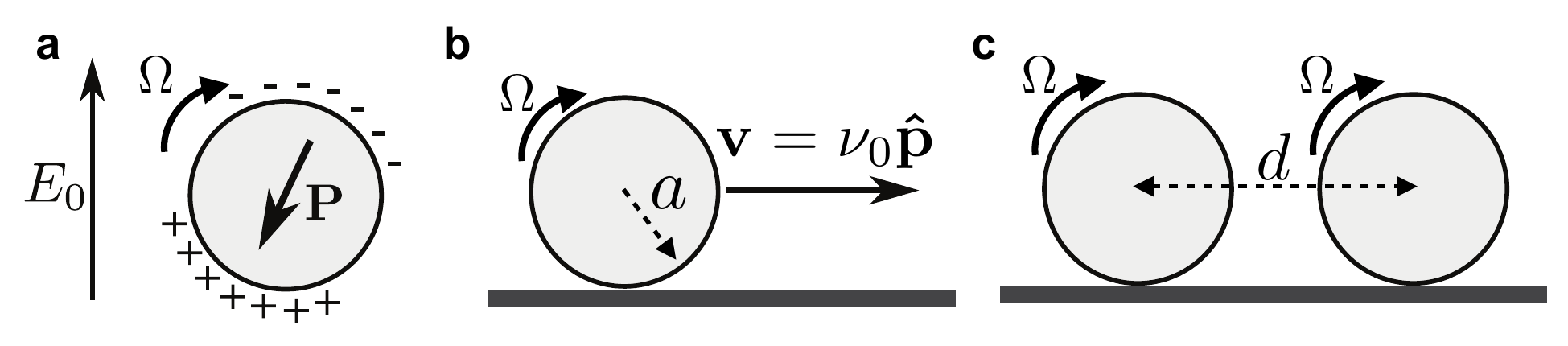}
		\caption{\textbf{Quincke rollers.} \textbf{a.} When  applying a DC electric field $\mathbf{E_0}$ to an insulating sphere immersed in a conducting fluid, a charge dipole forms at the sphere surface. When $E_0>E_{\rm Q}$, the electric dipole makes a finite angle with the electric field causing the steady rotation of the sphere at constant angular speed $\Omega$.  \textbf{b.} The rotation is converted into translation by letting the sphere to sediment on one electrode. When isolated, the resulting Quincke rotor rolls without sliding at constant speed: $\nu_0(0)=a\Omega$. \textbf{c.} When two colloids rolling in the same direction are close to each other the lubrication toque acting on the two spheres  separated by a distance d scales as $\ln d$ and hinders their rolling motion.} 
		\label{FigSupp1}
		\end{center}
\end{figure}

We conjecture a possible microscopic mechanism to explain the arrest of the Quincke rotation at high area fraction: the frustration of rolling motion by lubrication interactions, Fig.~\ref{FigSupp1}c. 
The viscous torque  $\mathbf T_{\rm V}$ acting on two nearby colloids rolling along the same direction is chiefly set by the lubricated flow in the contact region separating the two spheres.
$T_{\rm V}$ therefore increases logarithmically with $d-2a$ where $d$ in the interparticle distance~\cite{KimKarrila}. As there exists an upper bound to the magnitude of the electric torque $T_{\rm E}$, torque balance requires the rolling motion to become vanishingly slow as  $d-2a$ goes to zero: lubricated contacts frustrate collective motion.

\section{Experimental and Numerical Methods}
\label{Method}
\subsection{Experiments}
The experimental setup is identical to that described in~\cite{Geyer2018}.
We disperse  polystyrene colloids of radius $a=2.4\,\rm\mu m$ (Thermo Scientific G0500) in a solution of hexadecane including  $0.055\,\rm wt\%$ of AOT salt. We inject the solution in microfluidic chambers made of two electrodes spaced by a $110\,\rm \mu m$-thick scotch tape. The electrodes are glass slides, coated with indium tin oxide (Solems, ITOSOL30, thickness: $80\,\rm nm$). We apply a DC electric field between the two electrodes ranging from $1.3\,\rm V/\mu m$ to $2.6\,\rm V/\mu m$ using a voltage amplifier. If not specified otherwise, the data correspond to experiments performed at $1.8\,\rm V/\mu m$, i.e. at $E_0/E_{\rm Q}=2$.
We confine the  rollers inside  racetrack by coating the bottom electrode with an insulating pattern preventing the electric current to flow outside of the racetrack. To do so, we apply  $2\,\rm \mu m$-thick layer of insulating photoresist resin (Microposit S1818) and pattern it by means of conventional UV-Lithography as explained in~\cite{Geyer2018}. 

 In order to  keep track of the individual-colloid position and velocity, we image the system with a Nikon AZ100 microscope with a 4.8X magnification and record videos with a CMOS camera (Basler Ace) at framerate up to $900\,\rm Hz$. We use conventional techniques to detect and track all particles~\cite{Grier,Lu2007,Blair}. When performing large-scale observations we use a different setup composed of a $60\, \rm{mm}$ macro lens (Nikkor f/2.8G, Nikon) mounted on a $8\, \rm{Mpxls}$, 14bit CCD camera  (Prosilica GX3300).
\subsection{Numerics}
The numerical resolution of the Toner-Tu equations Eqs.~\eqref{Toner-1} and \eqref{Toner-2} were done using a semi-spectral method with semi-implicit Euler scheme.

\section{Linear Stability of the generalized Toner-Tu equations}
\label{app:linstab}
\FloatBarrier

In this appendix we show how the succession of instabilities of the homogeneous solutions of Eqs.~\eqref{Toner-1} and~\eqref{Toner-2} correctly predict the full phase behavior observed in our experiments and numerical simulations. 
\subsection{Stability of the disordered gas}
\label{app:linstab:gas}
We start by considering a homogneous gas phase where $\rho=\rho_{0}$, $W=0$. The
linearized dynamics of a small perturbation $\delta X\equiv(\delta \rho, \delta W)$ is given,
in Fourier space, by $\delta\dot X_k=M_k\delta X_k$, where the dynamical matrix $M_k$ is given by
\begin{multline}
M_k =\begin{pmatrix}
-D_{\rho}k^{2} & -ik \\
-(\epsilon_{1}+\epsilon_{1}'\rho_{0})ik \; & (\rho_{0}\epsilon_{2}-\phi)- D_{W}k^{2}
\end{pmatrix}
\end{multline}
The eigenvalues $\lambda_\pm$  of $M_k$ determines the stability of the gas phase. We find:
 \begin{widetext}
\begin{align}
\lambda_{\pm}= \frac{-(D_{\rho}k^{2}+D_{W}k^{2}-\rho_{0}\epsilon_{2}+\phi)\pm\sqrt{(D_{\rho}k^{2}+D_{W}k^{2}-\rho_{0}\epsilon_{2}+\phi)^{2}-4k^{2}(\epsilon_{1}+\epsilon_{1}') -4(D_{W}k^{2}-\rho\epsilon_{2} + \phi)D_{\rho}k^{2}}}{2}
\end{align}
\end{widetext}
The gas is therefore unstable when either one of the following
condition is satisfied:
\begin{align}
\rho_{0}\epsilon_{2}-\phi &>0 \\
 (\epsilon_{1}+\rho_{0}\epsilon_{1}') &<- (D_{W}k^{2}-\rho\epsilon_{2} + \phi)D_{\rho} 
\end{align}
The first condition is the standard spinodal instability leading to the
emergence of collective motion and local orientational order~\cite{Marchetti_review,Baskaran2010}. Beyond this classical results, we find a second instability akin to the
MIPS criterion in the presence of translationnal
diffusion~\cite{Cates_Review}. This second condition arise from the decrease of the pressure term in Eq.~\eqref{Toner-2} when $\rho$ increases. This condition is not met in our experiments where we find that the emergence of polar order occurs before the onset of solidification.

\subsection{Stability of  polar liquids}
Let us now consider a polar liquid where $\rho=\rho_{0}$, and $W=W_{0}$ with
$a_{4}W_{0}^{2}=(\rho_{0}\epsilon_{2}-\phi)$. Following the same procedure as in the previous section, the linearized dynamics
of a small perturbation $\delta X=(\delta\rho,\delta W)$ is defined by the dynamical matrix:
\begin{align}
M_k= \begin{pmatrix}
M_{\rho\rho}&M_{\rho W}\\M_{W\rho}&M_{WW}
\end{pmatrix}
\end{align}
where
\begin{align}
M_{\rho\rho}&=
-D_{\rho}k^{2},\\
M_{\rho W}&= -ik, \\
M_{W\rho}&=
-(\epsilon_{1}+\epsilon_{1}'\rho_{0})ik + (\epsilon_{2}+\epsilon_{2}'\rho_{0})W_{0} \; \\
M_{WW}&= (\rho_{0}\epsilon_{2}-\phi)- D_{W}k^{2}-\lambda ikW_{0}-3a_{4}W_{0}^{2}
\end{align}

Looking for the condition under which an eigenvalue admits a positive
real part, so that the homogeneous solution becomes unstable, leads
to:
\begin{align}
\label{eq::instab}
-\alpha_{6}k^{6}-\alpha_{4}k^{4}-\alpha_{2}k^{2}-\alpha_{0} > 0
\end{align}
with
\begin{align}
\alpha_{6}&= D_{W}D_{\rho}(D_{\rho}+D_{W})/W_0^2 \\
\alpha_{4}&= \lambda^{2} D_{\rho}D_{W}+2D_{\rho}D_{W}(D_{\rho}+D_{W})2a_{4}W_{0}^{2}\nonumber \\
&+(D_{\rho}+D_{W})^{2}\left[D_{\rho}2a_{4}+(\epsilon_{1}+\epsilon_{1}'\rho_{0})/W_0^2\right] \\
\alpha_{2}&=\left[D_{\rho } 2a_{4}+\left[(D_{\rho }+D_W\right) \left(\rho_{0}  \epsilon _2'+\epsilon _2\right)\right]\nonumber\\ 
&+2 \left(D_{\rho }+D_W\right) 2a_{4} \left(D_{\rho } 2a_{4}W_{0}^{2} +  \left(\rho_{0}  \epsilon _1'+\epsilon _1\right)\right)\nonumber\\
&+D_{\rho } D_W \left(2a_{4}W_{0}\right){}^2-2 \lambda   D_{\rho } \left(\rho_{0}  \epsilon _2'+\epsilon _2\right) \\
\alpha_{0}&= \left(2a_{4}\right){}^2 \left[D_{\rho } 2a_{4}W_{0}^{2}+  \left(\rho_{0}  \epsilon _1'+\epsilon _1\right)\right]\nonumber\\
&+ \left(\rho_{0}  \epsilon _2'+\epsilon _2\right) 2a_{4}W_{0}^{2}-\left(\rho_{0}   \epsilon _2'+W_{0} \epsilon _2\right)^{2} 
\end{align}
The analysis of the polynomial equation Eq.\eqref{eq::instab} leads to a  cumbersome
instability criterion, that however simplifies in the context of our
experiments where $\phi \ll \bar \rho$. 

At low density, $\rho_{0}
\ll \bar{\rho}$, so that $\rho_{0} \epsilon _1'+\epsilon
_1=\sigma_{1}$ and $\rho_{0} \epsilon _2'+ \epsilon _2=\sigma_{2}$, and only the constant term $\alpha_{0}$ can be positive, which
yields a simplified instability criterion:
\begin{align}
&\sigma_{2}/(\sigma_{2}\rho_{0}-\phi) >8a_{4}D_{\rho}(\sigma_{2}\rho_{0}-\phi)+4a_{4}\sigma_{1}+2\lambda\sigma_{2} 
\end{align}
In the limit of small $D_{\rho}$, relevant for our
experiments~\cite{Geyer2018}, one recovers the spinodal instability criterion of a homogeneous polar liquid~\cite{Solon2015PRE}
\begin{align}
\label{eq:VicsekInst}
\phi+\frac{\sigma_{2}}{2\lambda\sigma_{2}+4a_{4}\sigma_{1}}\gtrsim\sigma_{2}\rho_{0}
\end{align}

For larger densities, deep in the polar-liquid phase when $\rho_{0} \sim \bar{\rho}$, the
instability criterion~\eqref{eq:VicsekInst} cannot be satisfied. However, $\epsilon_i'(\rho_0)$ is not
negligible anymore, and a novel instability dictates the dynamics. All the
$\alpha_0$, $\alpha_2$ and $\alpha_4$ terms can be  negative
when $\epsilon_{i}'\rho_{0}+\epsilon_{i}$ is sufficiently large and negative
(provided that $D_{W}>D_{\rho}$, which holds  in our experimental
conditions~\cite{Geyer2018}). Again, one recovers a standard MIPS
instability, which occurs in the vicinity of $\bar \rho$ provided that the
decrease of $\epsilon_i(\rho)$ is sharp enough. We note that $\alpha_0$
could also change and  become negative for
$\epsilon_{i}'\rho_{0}+\epsilon_{i}$ sufficiently large and
positive. This situation is however not relevant for our experimental system but hints towards
 a possible additional instability of the ordered polar liquid, which will be discussed elsewhere.

\section{Description of the Movies}
 
 \paragraph{Movie 1.} Movie 1 shows a polar liquid flowing along a microfluidic racetrack. Dimensions of the observation window: $1.3 mm \times 1.3 mm$. The movie is slowed down by a factor 3.8.
 
\paragraph{Movie 2.} Movie 2 Close-up in the microfluidic racetrack. We first see the polar liquid phase and then the compact active solid forming at one end and melting at the other end. Dimensions of the observation window: $1.3 mm \times 1.3 mm$. wide. The movie is slowed down by a factor 3.8.
 
 \paragraph{Movie 3.} Movie 3 shows a typical experiment in a racetrack where an active solid steadily propagate through the homogenous polar liquid it coexists with. The movie is sped up by a factor 30.

    \paragraph{Movie 4.} Movie 4 shows the coarsening dynamics of multiple active solids.  The movie is sped up by a factor 30.

      \paragraph{Movie 5.} Hysteresis dynamics upon cycling the magnitude of the electric-field. This movie correspond to the experiments of Figs. 3e and 3f in the main document. The movie is sped up by a factor 30.
        \paragraph{Movie 6.} Response to an electric-field quench The bulk of the active solid is destabilized at all scales. This phenomenon is reminiscent to a spinodal decomposition scenario. The movie is sped up by a factor 30.
        
        \paragraph{Movie 7.} Numerical simulations of Eqs (1-2) with the same parameters as in Fig 5a-c, but with $dx=0.1$ and $dt=0.01$ to access longer time scales. Cycling the density $\rho_0$ up and down shows a clear hysteresis loop. The polar liquid is stable up to a spinodal density $\rho^{s}_L$. After an initial instability, the coarsening process leads the system towards phase separation between a polar liquid at $\rho^{b}_{L}$ and a solid phase, which is propagating backward. When decreasing the density, the solid jam is seen down to $\rho_0\simeq \rho^b_L\ll \rho^s_L$ highlighting the hysteresis loop.

\bibliography{Biblio}

\end{document}